# The P versus NP Brief

**Type of paper:**
`Brief study`

**Date:**
[*]`September 2007`

**Correspondence:**
`Mikael Franzén`
`Willgood Institute`
`Vegagatan 58, V5 Unit 43`
`413 11 Göteborg`
`Sweden`

<u>`mikael.franzen@willgood.org`</u>

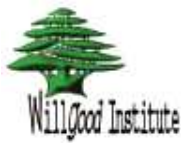

**Abstract:**
This paper discusses why P and NP are likely to be different. It analyses the essence of the concepts and points out that P and NP might be diverse by sheer definition. It also speculates that P and NP may be unequal due to natural laws.

---

[*] The paper was written and completed in Sept 2007 (RV01PNP01). It was then revised twice between the date of completion and April 2009, and hence two updated versions (RV02PNP01, RV03PNP01) were produced. As with RV02PNP01, changes in RV03PNP01 are minor and consist of additional content such as, notational declarations, formatting changes, and the latest correspondence details.

# 1. Introduction

It may be defined as a matter of polynomial time versus non-deterministic polynomial time; and hence the question whether P equals NP, or P does not equal NP. It is hailed the most significant problem within theoretical computer science and serves as a testimony of the complexity of seemingly simple logic and "everyday" axioms. But, it runs far deeper than that. In a sense it addresses the very nature of logic and physical truth; because the way we ask ourselves this question is well beyond the realm of mathematics, and therefore, it can be viewed as being metamathematical, as Scott Aronson[1] referred to it.

So, maybe the way of approach is far more important than the actual problem. Perhaps the manner in which P vs. NP is assessed and analyzed makes a significant difference in terms of importance. That is not to say that the essence of the problem is a variable depending on approach, but rather that the approach may hold the answer within itself.

Let us take a look at what the following entails:

P means Polynomial time and is defined as the computation runtime when *no greater* than the function of the problem. It would almost be foolish to assume that NP is anything else than the opposite of P. So NP means Non-deterministic Polynomial time, and is defined as the computation runtime when *greater* than the function of the problem. However, this definition is not exactly correct; the reason being indeterminism. In other words, problems in NP (besides those that are also in P) always rely on some sort of brute-force analysis or probabilistic measure. Therefore NP should be defined as the number of steps that may exponentially evolve in order to fit the relevant polynomial function. Thus, computation runtime is likely to be exponentially as long as a potential solution can be verified in polynomial time.[2]

Simply put, P problems are relatively fast to compute and hence also fast to solve and to verify, while NP problems are complicated to compute and therefore time-consuming to solve, but easy to verify once the answer is known. So, what is the differentiating factor in all this? Well, the variables of course. In order for P to equal NP, we need to eliminate the variables – Not literally speaking, but algorithmically speaking. However, is this at all possible?

# 2. Input, processor and output

Definition of (a), a polynomial time problem, and (b), a non deterministic polynomial time problem, where *k* is an independent constant; *n* the input string and *m* the instance of process work:

(a), $f(n) = O(k(n)) \rightarrow m(n) = O(n^k)$.

(b), $f(n) = O(k(n)) \rightarrow m(n) = O(n^k) \rightarrow m \neq n \rightarrow O(n(k)) \rightarrow m(n) = O(k^n)$.

In essence this tells us that a problem applied to the function *f(n)* characterizes that function by definition. Thus, a question in P or NP will render the function accordingly. So in general terms, all input with variables exceeding the number of functionalities of the processor, will most likely be NP type of problems. Let us examine this a little bit closer:

For example, providing an input of 1 and 1 to a processor, which performs addition, will result in an input of 2 units with an equal 2-unit processor functionality. The operation of addition will count as 1 unit and the production of a result, or output, may be considered as another 1 unit. Hence, in a straightforward manner, we could say that the plus (+) and equals (=) signs are the 2 units of functionality. This type of processor-analogy works very well for P problems, but it would not work for any type of NP problem. NP problems would not even fit the input intake;

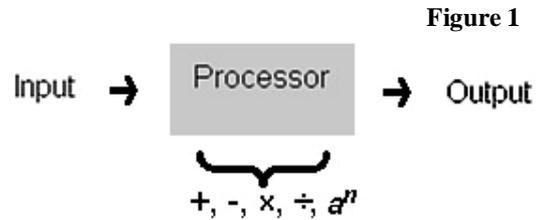

**Figure 1**

in a way it would be like feeding skittles through your inkjet, or putting stones in your juice blender. So, if we consider the Traveling salesman problem and the Subset sum problem: the variables are individually, or in some combination, processable by the kind of processor pictured above. However, all together it will render the processor in terms of [1]NTIME. So, concretely described, NP complexities are then solved by different, non-deterministic kinds of processors, as opposed to deterministic ones, like in Figure 1. However, non-deterministic processors are intrinsically dependent on deterministic processors (i.e. P is a subset of NP). Therefore, as implied earlier, a complexity function is characterized by its input, thus suggesting that P is not the same as NP; strongly implying, (perhaps naively), that in general, P is not equal to NP.

## 3. Characterizations

P complexities are processed at their very essence, instantly, even if time-wise it may take very long:

$$\forall n. \, m \propto f(n) = O(k(n)) \propto \forall m. \, n \leq O(n^k).$$

-Therefore, $n$ may approach infinity, and consequently, the function and its output approaches equal bounds. Although it possibly could be impractical and lingering, the runtime, on the other hand, does not approach infinity – Well, the total runtime may approach infinity to the extent that the input does, but, total runtime would normally not be counted until an end is reached, and subsequently, in this context, total runtime may be an invalid notion:[3]

$$^{II} \, f(n), \, m(n) > 0 \rightarrow \lim_{n \to \infty} \frac{m(n)}{f(n)} \rightarrow \sum_{n=1}^{\infty} f(n) \leftrightarrow \sum_{n=1}^{\infty} m(n).$$

NP complexities may or may not be processed at their essence. That is to say, a brute-force analysis, or some other type of non-deterministic activity may yield computational results that are relevant or irrelevant to the core of the question. Consequently:

$$^{III} \, \exists n. \, m \propto f(n) = O(k(n)) \propto \Omega \geq O(k^n).$$

---

[I] $O(f(n)) = NP = NTIME$, $NP = \bigcup_{k \in N} NTIME(n^k) \mid O(2^{f(n)}) = NEXPTIME = \bigcup_{k \in N} NTIME(2^{n^k})$

[II] Let the sequential terms $\{a_n\}$ and $\{b_n\}$ be the instances of functionality and input, $f(n)$ and $m(n)$

[III] Let $\Omega$ denote all possible outcomes, $\{x_1, x_2, ...\}$

-So, as with P complexities, *n* may approach infinity, but with one very significant difference, namely, the relationship between runtime and function. Here we see a dislocation between input, function and runtime. Firstly in terms of components and functionality; input may be non-deterministic in a way that yields no immediate function, or, it may be deterministic, but 'incorrect', thus yielding an irrelevant function. Secondly in terms of totality; input may produce disproportionate exponential or factorial runtimes, and runtime may depend on variables that are outside immediate and closely related sets. In other words, the scope of Ω may be considered unbounded by its own limits:

$$f(n), m(n) > 0 \rightarrow \Omega \vdash \lim_{n \to \infty} \frac{m(n)}{f(n)} \therefore \left(\frac{m(n)}{f(n)}\right)^{\Omega} \rightarrow \left(\sum_{n=1}^{\infty} f(n) \leftrightarrow \sum_{n=1}^{\infty} m(n)\right)^{\Omega}.$$

So:

If we create an algorithm that renders P to equal NP it would have to remodel all dependencies and variables into a nice polynomial package. In addition it needs a real-time performance by the square root of all possible answers:

$$f(n), m(n) > 0 = \sqrt{\Omega}$$ –This would yield the same expression as the one for P complexities on previous page, but the problem with this is that it first entails the following:

$$\sqrt{\lim_{n \to \infty} \left(\frac{m(n)}{f(n)}\right)^{\Omega} \rightarrow \left(\sum_{n=1}^{\infty} f(n) \leftrightarrow \sum_{n=1}^{\infty} m(n)\right)^{\Omega}} \propto \Omega$$ –And thus we are back where we started:

$$f(n), m(n) > 0 \rightarrow \Omega \vdash \lim_{n \to \infty} \frac{m(n)}{f(n)} \therefore \left(\frac{m(n)}{f(n)}\right)^{\Omega} \rightarrow \left(\sum_{n=1}^{\infty} f(n) \leftrightarrow \sum_{n=1}^{\infty} m(n)\right)^{\Omega}.$$

So in this context:
$$\sqrt{\Omega} = \Omega \because \sqrt{\Omega} \in \Omega.$$

Hence, trying to reduce size and limit the scope, while at the same time taking all the variables into account, results in an extrapolation of all possible outcomes (Ω). Therefore all steps that would normally be taken, when not using the algorithm, are reproduced as algorithmic processes that are a polynomial function of the input size.

## 5. Quantum connection?

Quantum mechanics deals with things at the very essence of reality. And in a mathematical sense it would not be farfetched to say the same thing about the P vs. NP problem. Consider for a moment the Heisenberg uncertainty principle… Probability rules; things are dealt with in a non-deterministic manner.

**What about Quantum computers and the equality of P and NP?**
To answer this question, let us go back to the latter: Nature at its core, as far as we know, is probabilistic; this indeterminism leads to the deterministic phenomena observable in our everyday macroscopic world. So does this not in itself answer the question? Does this mean that quantum computers, as a concept and a tool to solve the P vs. NP problem, are no better than "normal" computers? It probably does.

## 6. Conclusion

This brief has shown how NP type problems are intrinsically tied into themselves, almost like a subset concurrently outside and within the boundaries of a greater set. P problems, on the other hand, are free from such ties, and tend not to become bogged down within. So, it is tempting to speculate that if P equaled NP, this fact would be evident and clear. But of course, there may exist obscure conditions such that P equals NP; and the reason we cannot identify them could lie in possible short-comings of current mathematics. Hence we may need to discover new mathematics and/or a novel type of reasoning. However, given what has been presented, it must be said that the likelihood of P being equal to NP is very slim. It might just be that the laws of nature prohibit this equality: Think quantum mechanics, think indeterminism…–But, if drawing parallels to nature is valid, does not that mean that P is just as probabilistic as NP? Well, that is a fair point. However, remembering that quantum mechanical phenomena leads to classical phenomena, then NP complexities would be the mathematical counterpart of quantum mechanical processes or systems; subsequently, P complexities would be the mathematical counterpart of processes or systems in the classical world. Then, disregarding potential fallacies, P is to Newton as NP is to Heisenberg.

So, to end on a positive note: whatever our convictions; let us keep an open mind.

**References**

[1] Aaronson S, (2003), 'Is P Versus NP Formally Independent', *Bulletin of the EATCS 81*, p 1.

[2] Howe D (1995) 'Foldoc - *NTIME*' [on-line], London, UK, Imperial College. Available from: http://burks.bton.ac.uk/burks/foldoc/82/80.htm

[3] Weisstein, E W. 'MathWorld - *Limit Comparison Test*' [on-line] Champaign, USA, Wolfram Research Inc, Available from: http://mathworld.wolfram.com/LimitComparisonTest.html [accessed Aug 2007].

# Bibliography

Aaronson S, (2003), 'Is P Versus NP Formally Independent', *Bulletin of the EATCS 81*. Also available online from: http://www.scottaaronson.com/papers/ [accessed Aug 2007].

Broadhurst D -et al., (2002), *The Quantum World*, Milton Keynes, UK, The Open University, pp 26-41.

Howe D (1985-2007) 'Free On-line Dictionary of Computing (Foldoc)' [on-line], London, UK, Imperial College. Available from: http://foldoc.org/ [accessed July 2007].

GmH -et al., 'P versus NP' [on-line], New York, USA. Available from QEden: http://www.qeden.com/wiki/P_versus_NP:Math_Intense_Intro [accessed Aug 2007].

Neill H (2003), *Calculus*, 3rd edtn, London, UK, Hodder Headline ltd - tech yourself, pp 113-123.

Nielsen D -et al., (2002), *Dictionary of Mathematics*, 3rd edtn, London, UK, Penguin Books, pp 254, 301.

Weisstein, E W. 'Complexity Theory' [on-line] Champaign, USA. From MathWorld –A Wolfram Web Resource: http://mathworld.wolfram.com/ComplexityTheory.html [accessed July 2007].